\documentclass{PoS}

\def\beq{\begin{equation}}
\def\eeq{\end{equation}}
\def\bea{\begin{eqnarray}}
\def\eea{\end{eqnarray}}

\def\O{y}

\def\d{\hbox{d}}

\def\ln{\hbox{ln}}

\title{Event shapes in $e^+e^-$ annhilation at NNLO}

\ShortTitle{$e^+e^-$ event shapes at NNLO}

\author{A.~Gehrmann--De Ridder\\
Institute for Theoretical Physics, ETH, CH-8093 Z\"urich,
Switzerland}

\author{\speaker{T.~Gehrmann}
\thanks{Supported in part by the Swiss National Science Foundation
(SNF) under contract 200020-117602.}\\
Institut f\"ur Theoretische Physik, Universit\"at Z\"urich,
CH-8057 Z\"urich, Switzerland}

\author{E.W.N.~Glover\\
Institute of Particle Physics Phenomenology, 
        Department of Physics,\\
        University of Durham, Durham, DH1 3LE, UK}

\author{G.\ Heinrich
\thanks{Supported by the UK Science and Technology Facilities Council.}
\\
Institute of Particle Physics Phenomenology, 
        Department of Physics,\\
        University of Durham, Durham, DH1 3LE, UK}

\abstract{We report first results on the calculation of NNLO 
corrections to event shape distributions in electron-positron annhilation. 
The corrections are sizeable for all
variables,  however their magnitude is substantially different for different
observables. We observe that 
inclusion of the  NNLO corrections yields a considerably 
better agreement between theory and experimental data both in shape and 
normalisation of the event shape distributions.}

\FullConference{8th International Symposium on Radiative Corrections (RADCOR)\\
		 October 1-5 2007\\
		 Florence, Italy}

\begin{document}

\section{Introduction}

For more than a decade experiments at LEP (CERN) and SLC (SLAC) 
gathered  a wealth of high precision high energy hadronic data
from electron-positron annihilation at a range of centre-of-mass 
energies~\cite{ALEPH-qcdpaper,lep,sld}. 
This data provides one of the    
 cleanest
ways of probing our quantitative understanding of QCD. 
This is particularly so because the strong interactions occur only in 
the final state and are not entangled with the parton density functions associated 
with beams of hadrons.
As the understanding of the strong interaction, and the capability of 
making more precise theoretical predictions, develops, 
more and more stringent comparisons of theory and experiment are possible,
leading to improved measurements
of fundamental quantities such as the strong 
coupling constant~\cite{expreview}.

In addition
to measuring multi-jet production rates, more specific information  about the
topology of the events can be extracted. To this end, many variables  have been
introduced which characterise the hadronic structure of an event. 
With the precision data from LEP and SLC, experimental
distributions for such event shape variables have been extensively  studied and
have been compared with theoretical calculations based on next-to-leading order
(NLO)  parton-level event generator  programs~\cite{ERT,kunszt,event}, 
  improved by
resumming kinematically-dominant leading and next-to-leading logarithms
(NLO+NLL)~\cite{ctwt}  and by the inclusion of  
non-perturbative models of power-suppressed hadronisation
effects~\cite{power}. 

The precision of the strong coupling constant 
determined from event shape data has been limited up to now 
largely by the scale
uncertainty of the perturbative NLO calculation. We report here on the 
first calculation of NNLO corrections to event shape variables, and discuss 
their phenomenological impact.

\section{Event shape variables}
\label{sec:shapes}

In order to characterise hadronic final states in electron-positron
annihilation, a variety of event shape variables have been proposed in 
the literature, for a review see e.g.~\cite{QCDbooks}. These variables can be categorised 
into different classes, 
according to the minimal number of final-state particles required for them 
to be non-vanishing: In the following we shall only consider three particle final states which are thus closely related to three-jet final states.

Among those shape variables,
six variables~\cite{shapes}
 were studied in great detail: the thrust $T$, the
normalised heavy jet mass $\rho$, 
the wide and total jet
broadenings $B_W$ and $B_T$,  
the $C$-parameter and the transition from three-jet to 
two-jet final states in the Durham jet algorithm $Y_3$.

The perturbative expansion for the distribution of a 
generic observable $\O$ up to NNLO at $e^+e^-$ centre-of-mass energy $\sqrt{s}$, 
for a renormalisation scale $\mu^2$  is given by
\begin{eqnarray}
\frac{1}{\sigma_{{\rm had}}}\, \frac{\d\sigma}{\d y} (s,\mu^2,y) &=& 
\left(\frac{\alpha_s{}(\mu^2)}{2\pi}\right) \frac{\d \bar A}{\d y} +
\left(\frac{\alpha_s{}(\mu^2)}{2\pi}\right)^2 \left( 
\frac{\d \bar B}{\d y} + \frac{\d \bar A}{\d y} \beta_0 
\log\frac{\mu^2}{s} \right)
\nonumber \\ &&
+ \left(\frac{\alpha_s{}(\mu^2)}{2\pi}\right)^3 
\bigg(\frac{\d \bar C}{\d y} + 2 \frac{\d \bar B}{\d y}
 \beta_0\log\frac{\mu^2}{s}
\nonumber \\ &&
\hspace{24mm} + \frac{\d \bar A}{\d y} \left( \beta_0^2\,\log^2\frac{\mu^2}{s}
+ \beta_1\, \log\frac{\mu^2}{s}   \right)\bigg)+ {\cal O}(\alpha_s{4})  \;.
\label{eq:NNLOmu} 
\end{eqnarray}
The dimensionless 
perturbative coefficients $\bar A$, $\bar B$ and $\bar C$ depend only 
on the event shape variable $y$. They are computed by a fixed-order 
parton-level calculation, which includes final states with three partons 
at LO, up to four partons at NLO and up to five partons at NNLO. 
LO and NLO corrections to event shapes have been available already for 
a long time~\cite{ERT,kunszt,event}. 

 The calculation of the NNLO corrections is carried out using 
a newly developed
parton-level event generator programme {\tt EERAD3} which contains 
the relevant 
matrix elements with up to five external partons~\cite{3jme,muw2,V4p,tree5p}. 
Besides explicit infrared divergences from the loop integrals, the 
four-parton and five-parton contributions yield infrared divergent 
contributions if one or two of the final state partons become collinear or 
soft. In order to extract these infrared divergences and combine them with 
the virtual corrections, the antenna subtraction method~\cite{ant} 
was extended to NNLO level~\cite{ourant} and implemented
for $e^+e^- \to 3\,\mathrm{jets}$ and related event-shape variables~\cite{eerad3}. The analytical cancellation of all 
infrared divergences serves as a very strong check on the implementation. 
{\tt EERAD3} yields the perturbative  $A$, $B$ and $C$ coefficients as 
histograms for all infrared-safe event-shape variables 
related to three-particle 
final states at leading order. From these, 
 $\bar A$, $\bar B$ and $\bar C$ are computed by normalising to the total 
hadronic cross section.
As a cross check, the $A$ and $B$  coefficients have also been obtained from an independent integration~\cite{event}
of the NLO matrix elements~\cite{ERT}, showing excellent agreement. 

For small values of the event shape variable $y$, the fixed-order expansion, 
eq.\ (\ref{eq:NNLOmu}), fails to converge, 
because the fixed-order coefficients are enhanced by powers of $\ln(1/y)$.
In order to obtain reliable predictions
in the region of $y \ll 1$ it is necessary to resum entire sets of logarithmic terms at all orders in $\alpha_s$. 
A detailed description of the predictions at next-to-leading-logarithmic approximation (NLLA) can
be found in Ref.\ \cite{as_theory-uncertainties}.

\section{NNLO results}

The precise size and shape of the NNLO corrections depend on the observable 
in question. Common to all observables is the divergent behaviour of 
the fixed-order prediction in the two-jet limit, where soft-gluon effects 
at all orders become important, and where resummation is needed. For several 
event shape variables 
 (especially $T$ and $C$) the full kinematical range is not yet realised 
for three partons, but attained only in the multi-jet limit. In this case,
the fixed-order description is also insufficient since it is  limited 
to a fixed multiplicity (five partons at NNLO). Consequently, the 
fixed-order description is expected to be reliable in a restricted 
interval bounded by the two-jet limit on one side and the multi-jet 
limit on the other side. 

\begin{figure}[t]
\begin{center}
\includegraphics[angle=-90,width=10cm]{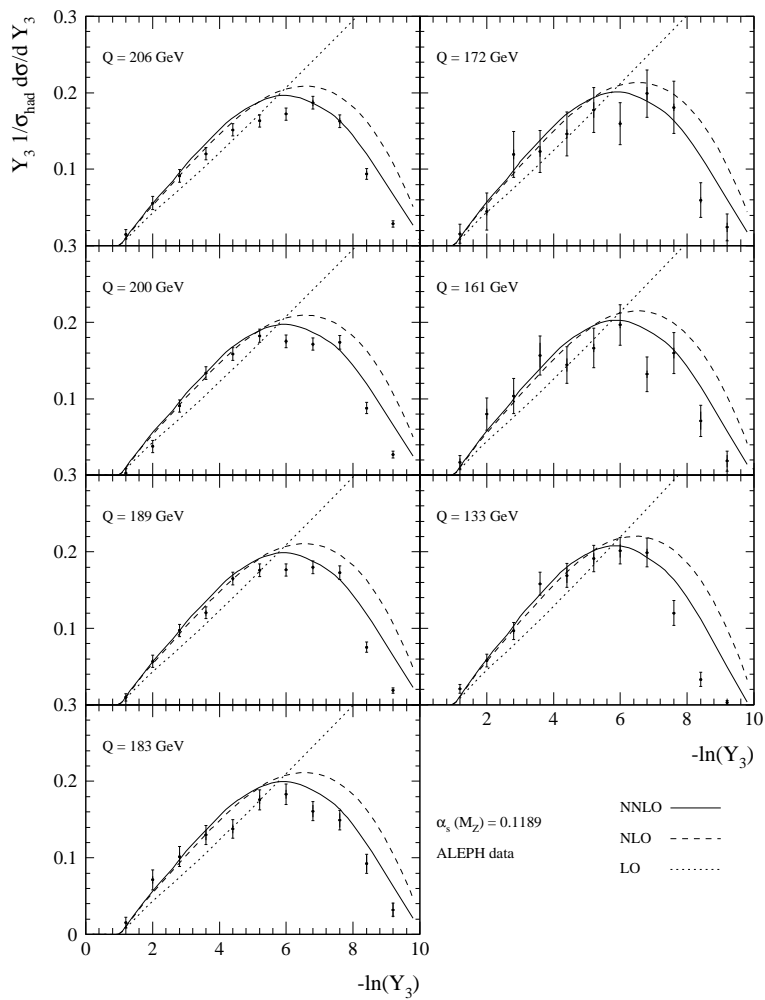}
\end{center}
\caption{\small Perturbative fixed-order predictions 
for the $Y_3$-distribution, compared to LEP2 data from ALEPH.} 
\protect\label{fig:y23}
\end{figure}

In this intermediate region, we observe that 
inclusion of  NNLO corrections (evaluated at the $Z$-boson mass, and 
for fixed value of the strong coupling constant) typically increases 
the previously available NLO prediction. 
The magnitude of this increase differs considerably between 
different observables\cite{ourevent}, 
it is substantial for $T$ (18\%), $B_T$ (17\%)  and 
$C$ (15\%), moderate for $\rho$ and $B_W$ (both 10\%) and small for 
$Y_3$ (6\%). For all shape variables, we observe that the renormalisation
scale uncertainty of the NNLO prediction is reduced by a factor 2 or more
compared to the NLO prediction. 
Inclusion of the NNLO corrections modifies the shape of the event shape 
distributions. We observe that 
the NNLO prediction describes the shape of the measured event shape 
distributions over a wider kinematical range than the NLO prediction, both 
towards the two-jet and the multi-jet limit. To illustrate the 
impact of the NNLO corrections, we compare the fixed-order predictions 
for $Y_3$ to LEP2-data obtained by the ALPEH experiment in 
Figure~\ref{fig:y23}, which illustrates especially the improvement
in the approach to the two-jet region (large $-\ln(Y_3)$).   

The information contained in the event shape distributions can be 
restructured by computing individual moments. Moments of event shape 
distributions have been studied 
theoretically and experimentally in particular in view of 
understanding non-perturbative power corrections~\cite{power}.
Consequently, perturbative NNLO corrections will improve the discrimination 
between higher perturbative orders and genuine non-perturbative effects. 
For the first moment $\langle 1-T \rangle$ of the thrust distribution, we find
the integrated coefficients
\begin{displaymath}
{\cal A} = 2.101\,\qquad {\cal B} = 44.98\,\qquad 
{\cal C} = 1095 \pm 130\;,
\end{displaymath}
which yields for $\sqrt{s}=\mu=M_Z$:
\begin{displaymath}
\langle 1-T \rangle(\alpha_s(M_Z) = 0.1189) 
= 0.0398\, ({\rm LO})\; +\; 0.0146\, ({\rm NLO})  \; 
+ \; 0.0068 \, ({\rm NNLO})\;.
\end{displaymath}
Work on moments of the event shapes is ongoing.

\section{Determination of the strong coupling constant}
Using the newly computed NNLO corrections to event shape variables, we
performed\cite{ouras} 
a new extraction of $\alpha_s$ from data on the standard set of 
six event shape variables, measured 
 by the ALEPH\ collaboration \cite{ALEPH-qcdpaper}
at centre-of-mass energies of 91.2, 133, 161, 172, 183, 189, 200 and 206 GeV.
The combination of 
all NNLO determinations from all shape variables  yields 
\begin{displaymath}
    \alpha_s(M_Z) = 0.1240 \;\pm\; 0.0008\,\mathrm{(stat)}
     					 \;\pm\; 0.0010\,\mathrm{(exp)}
                                   \;\pm\; 0.0011\,\mathrm{(had)}
                                   \;\pm\; 0.0029\,\mathrm{(theo)} .
 \end{displaymath}
We observe a clear improvement in the fit quality when going to
NNLO accuracy. Compared to NLO the value of $\alpha_s$ is lowered 
by about 10\%, but still higher than for NLO+NLLA~\cite{ALEPH-qcdpaper},
 which 
shows the obvious need for a matching of NNLO+NLLA for a fully reliable 
result.  
 The scatter among the
 $\alpha_s$-values extracted from different shape variables is 
lowered considerably, and the theoretical uncertainty is decreased by 
a factor 2 (1.3) compared to NLO (NLO+NNLA). 

These observations visibly illustrate the improvements gained from 
the inclusion of the NNLO corrections, and highlight the need for 
further studies on the matching of NNLO+NLLA, and on the 
derivation of NNLLA resummation terms.

\section{Outlook}
Our results for the NNLO corrections open up a whole 
new range of possible 
comparisons with the LEP data.
The potential of these studies is
illustrated by the new determination of 
$\alpha_s$ reported here, which can be 
further improved by the matching NLLA+NNLO, currently in progress. 
Similarly, our results will also allow a renewed study of
power corrections, now matched to NNLO.


\bibliographystyle{JHEP}

\end{document}